\documentclass[]{spie}  
 
\usepackage{amsmath,amsfonts,amssymb,aas_macros}
\usepackage{graphicx}
\usepackage[table,x11names]{xcolor}
\usepackage[square,sort,comma,numbers]{natbib}
\usepackage[colorlinks=true, allcolors=blue]{hyperref}
\usepackage{placeins}
\usepackage{subcaption}

\title{The Blue Multi Unit Spectroscopic Explorer (BlueMUSE) on the VLT: science drivers and overview of instrument design}

\author[a]{Johan Richard}
\author[a]{Rémi Giroud}
\author[a]{Florence Laurent}
\author[b]{Davor Krajnovi\'c}
\author[a]{Alexandre Jeanneau}
\author[a]{Roland Bacon}
\author[c]{Manuel Abreu}
\author[d]{Angela Adamo}
\author[e]{Ricardo Araujo}
\author[a]{Nicolas Bouché}
\author[c]{Jarle Brinchmann}
\author[f]{Zhemin Cai}
\author[b,m]{Norberto Castro}
\author[g]{Ariadna Calcines}
\author[e]{Diane Chapuis}
\author[d]{Adélaïde Claeyssens}
\author[o]{Luca Cortese}
\author[h]{Emanuele Daddi}
\author[g]{Christopher Davison}
\author[f]{Michael Goodwin}
\author[g]{Robert Harris}
\author[d]{Matthew Hayes}
\author[j]{Mathilde Jauzac}
\author[b]{Andreas Kelz}
\author[e]{Jean-Paul Kneib}
\author[k]{Audrey A. Lanotte}
\author[f]{Jon Lawrence}
\author[h]{Vianney Le Bouteiller}
\author[i]{Rémy Le Breton}
\author[a]{Matthew Lehnert}
\author[f]{Angel Lopez Sanchez}
\author[f]{Helen McGregor}
\author[j]{Anna F. McLeod}
\author[c]{Manuel Monteiro}
\author[g]{Simon Morris}
\author[p]{Cyrielle Opitom}
\author[a]{Arlette Pécontal}
\author[f]{David Robertson}
\author[b]{Matin M. Roth}
\author[n]{Jesse van de Sande}
\author[j]{Russell Smith}
\author[b]{Matthias Steinmetz}
\author[j]{Mark Swinbank}
\author[b]{Tanya Urrutia}
\author[k]{Anne Verhamme}
\author[b]{Peter M. Weilbacher}
\author[l]{Martin Wendt}
\author[k]{François Wildi}
\author[f]{Jessica Zheng}
\author[a-o]{the BlueMUSE consortium}

\affil[a]{Université Lyon 1, CNRS, Centre de Recherche Astrophysique de Lyon (CRAL), Saint-Genis-Laval, France}
\affil[b]{Leibniz-Institut für Astrophysik Potsdam (AIP), Germany}
\affil[c]{Instituto de Astrofísica e Ciências do Espaço, Lisboa, Portugal}
\affil[d]{Stockholm University, Stockholm, Sweden}
\affil[e]{Ecole Polytechnique Fédérale de Lausanne (EPFL), Lausanne, Switzerland}
\affil[f]{Australian Astronomical Optics, Macquarie Univ., Sydney, Australia}
\affil[g]{Ctr. for Advanced Instrumentation, Durham Univ., Durham, United Kingdom}
\affil[h]{Universit\'{e} Paris-Saclay, Universit\'{e} Paris Cit\'{e}, CEA, CNRS, AIM,  Gif-sur-Yvette, France}
\affil[i]{Université Paris-Saclay, CEA, IRFU, Gif-sur-Yvette, France}
\affil[j]{Centre for Extragalactic Astronomy, Department of Physics, Durham University, Durham, UK}
\affil[k]{Dpt d’Astronomie, Univ. de Genève, Versoix, Switzerland}
\affil[l]{Univ Potsdam, Inst Phys \& Astron, Potsdam, Germany}
\affil[m]{Institut für Astrophysik und Geophysik, Göttingen, Germany}
\affil[n]{School of Physics, University of New South Wales \& ARC Centre of Excellence for All Sky Astrophysics in 3 Dimensions (ASTRO 3D), Australia}
\affil[o]{International Centre for Radio Astronomy Research, The University of Western Australia, Australia}
\affil[p]{Institute for Astronomy, University of Edinburgh, Royal Observatory, Edinburgh, UK}

\authorinfo{Further author information: \\Johan Richard: E-mail: johan.richard@univ-lyon1.fr}

\begin{document} 
\maketitle

\begin{abstract}
BlueMUSE is a blue-optimised, medium spectral resolution, panoramic integral field spectrograph under development for the Very Large Telescope (VLT). With an optimised transmission down to 350 nm, spectral resolution of R$\sim$3500 on average across the wavelength range, and a large FoV (1 arcmin$^2$), BlueMUSE will open up a new range of galactic and extragalactic science cases facilitated by its specific capabilities. The BlueMUSE consortium includes 9 institutes located in 7 countries and is led by the Centre de Recherche Astrophysique de Lyon (CRAL). The BlueMUSE project development is currently in Phase A, with an expected first light at the VLT in 2031. We introduce here the Top Level Requirements (TLRs) derived from the main science cases, and then present an overview of the BlueMUSE system and its subsystems fulfilling these TLRs. We specifically emphasize the tradeoffs that are made and the key distinctions compared to the MUSE instrument, upon which the system architecture is built.

\end{abstract}

\keywords{BlueMUSE, Very Large Telescope, Integral field spectroscopy, Optical Design, Blue / UV spectral range}

\section{INTRODUCTION}
\label{sec:intro}  

The advent of Integral Field Spectroscopy (IFS) has revolutionised astronomical instrumentation by enabling to perform the spectroscopy of \textit{everything} within a given field of view. One of the most recent improvements in the technique is the concept of wide-field, panoramic IFS, which pushes even further the multiplexing capabilities, allowing us to observe even more extended sources, as well as perform large and blind surveys of thousand of sources over a contiguous sky region. 

The Multi Unit Spectroscopic Explorer (MUSE) is one of the second-generation instruments of the ESO Very Large Telescope (VLT) and the largest panoramic IFS (1 arcmin$^2$) currently in operation on a 8-10m class telescope. Coupled with the adaptive optics system, MUSE image quality and throughput  have made it a tremendous success in the astronomical community \footnote{this is clearly illustrated by ESO Telbib statistics \url{https://www.eso.org/sci/php/libraries/telbibstats/}}

The Blue Multi-Unit Spectroscopic Explorer (BlueMUSE) has been selected by ESO as a new panoramic 1 arcmin$^2$ IFS on the VLT as part of the VLT2030 instrument suite. BlueMUSE aims at extending the current limits of MUSE in wavelength and spectral resolution by covering the near-UV and blue wavelength range (down to 350 nm) at even higher spectral resolution (average R$\sim$3500). 

\bigskip\par

\textbf{BlueMUSE Consortium and Project timeline}

\begin{figure}[ht]
    \centering
    \includegraphics[width=1\linewidth]{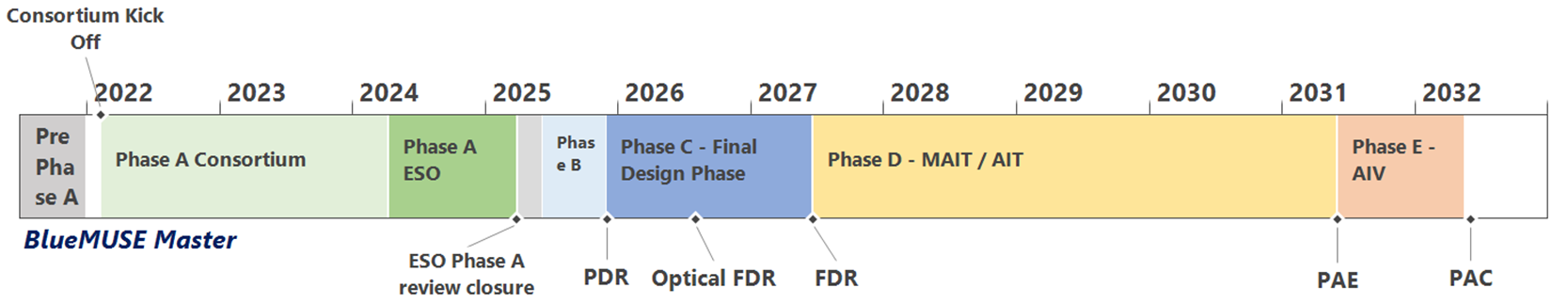}
    \caption{BlueMUSE Master Schedule}
    \label{fig:schedule}
\end{figure}
\bigskip\par

BlueMUSE builds upon the heritage of MUSE and follows a similar system architecture, but includes both obvious and novel improvements (e.g., better temperature control and alignment processes). The current BlueMUSE design includes an overall simplification in the number of spectrographs (16) while covering the same field of view. The BlueMUSE project is currently starting its Phase A, with an expected first light at the telescope in 2031 (Fig.~\ref{fig:schedule}). The project involves 9 institutes in addition to ESO, with their respective contribution presented in Table~\ref{tab:wp}.

\begin{table}[ht!]
    \footnotesize
    \centering
\begin{tabular}{|l|c|}
         \hline&\\
         {\bf Institute  / Country} & {\bf Phase A Work-Package} \\[2ex]
         \hline
         
         \hline
         CRAL (FR) &
         \begin{tabular}{@{}c@{}}Management \\
         Optics global design \& system engineering \\
         Slicer design
         \end{tabular} 
          \\
         \hline
         CEA-IRFU and AIM (FR) &
         \begin{tabular}{@{}c@{}}Detector Vessel \\
         Vacuum and Cryogenic System \\
         \end{tabular} \\
         \hline
         Astralis-AAO-MQ (AUS) &
         \begin{tabular}{@{}c@{}} Mechanical global design \& system engineering \\
         Splitting and Relay Optics \\ Main Structure and Enclosure
         \end{tabular} \\
         \hline
         AIP/UP/IAG (DE) &
         \begin{tabular}{@{}c@{}} Calibration Unit design \\
         Data reduction software \\
         End-to-end science simulator (BlueSi)
         \end{tabular} \\
         \hline
         IA/CAUP (PT) &
         \begin{tabular}{@{}c@{}} Instrument Control Software \\
         Instrument Control Electronics
         \end{tabular} \\
         \hline
         EPFL (CH) &
         \begin{tabular}{@{}c@{}} Automated Relay Optics Alignment \\
         \end{tabular} \\[2ex]
         \hline
         Geneva University (CH) &
         \begin{tabular}{@{}c@{}} Fore-Optics design \\
         \end{tabular} \\[2ex]
         \hline
         Stockholm University (SE) &
         \begin{tabular}{@{}c@{}} Data Analysis Software \\
         \end{tabular}  \\[2ex]
         \hline
         CfAI \& Durham University (UK) &
         \begin{tabular}{@{}c@{}} Spectrograph design \\
         \end{tabular} \\[2ex]
         \hline
    \end{tabular}
    \caption{\label{tab:wp}Consortium institutes and Phase A WP allocation.}
\end{table}

In seven years from now, at a time when the focus of most of the new large facilities (ELT, JWST) will be on the infra-red, BlueMUSE will be a unique facility.
 It will clearly outperform any ELT instrument in the blue/UV. Its synergy with ELT and JWST is strong, but also with ALMA, SKA, \textit{Euclid} and \textit{Athena}.

We present in this manuscript some technical details on the BlueMUSE instrument and the science it will allow. Section \ref{sec:science} introduces some of the main science cases and the Top Level Requirements  derived from them. We then move to describing the system architecture design in Sect.~\ref{sec:architecture} and a first estimate on the performances from our preliminary optical design in Sect.~\ref{sec:perfs}.

\section{BlueMUSE Science Requirements}

\label{sec:science}
\subsection{Overview and key science cases} 

We briefly summarize below science cases for which the BlueMUSE instrument will provide unique capabilities, and which are constraining some of the Top Level Requirements (TLRs); they are distributed in three categories: Local Volume, Nearby Galaxies and Distant Universe. Although representative of the kind of science allowed by BlueMUSE, this list of science cases does not fully cover the full range of science allowed by such a generic instrument.

The main performance requirements derived from the science cases are summarized in Table \ref{tab:tlr}. We highlight one ‘key science case’ in each category (marked in blue) for which the compliance of the instrument TLRs is mandatory.


\subsubsection{Local Volume}

\textbf{Key Science Case: Wolf-Rayet (WR) and OB star evolution}


Massive stars are the main chemical and dynamic engines of the Universe and key to interpreting phenomena in high-redshift galaxies. Studying the early reionization epoch and formation of the first galaxies, the transition to population II metallicities, rates of very luminous supernovae, gamma-ray bursts, and gravitational waves requires, firstly, to understand the most massive stars born in the Universe. Nevertheless, the formation and evolution of very massive stars are still uncertain (\cite{Langer2012}). This lack of knowledge propagates to star formation studies in galaxies where individual stars remain unresolved, thus blurring the conclusions. It is mandatory to understand the evolution and the formation of the most massive stars in the Milky Way and Local Group before moving forward to more distant galaxies. 

Spectroscopy of massive stars in star-forming regions is notoriously difficult due to nebular contamination and crowding. Additionally, gathering large spectroscopic samples can be time-consuming and expensive in terms of observing time. BlueMUSE offers a novel and unique solution to study blue massive stars and overcome these challenges, building on the path established by MUSE (e.g., \cite{Kamann2018, Castro2018, Roth2018}). It will enhance the capabilities of MUSE by improving multiplexity, sensitivity, field-of-view, and spectral resolution. Since hot stars are intrinsically blue, BlueMUSE's wavelength coverage will be closer to the peak of their spectral energy distribution compared to red MUSE. Furthermore, the optical blue wavelength range explored with BlueMUSE will provide access to a rich set of transitions for elements such as silicon, magnesium, oxygen, carbon, and nitrogen. These elements are crucial for constraining stellar evolution and characterizing the stellar atmospheres of massive stars (e.g., \cite{Castro2012}).

\noindent
\textbf{Resolved HII regions, SNRs, and PNe}

HII regions in local galaxies are the signposts for massive main sequence stellar feedback. Supernovae remnants (SNRs) and planetary nebulae (PNe) probe feedback from dying stars.  Optical spectroscopy of these different systems is a powerful tool to trace both physical conditions of the ionised gas (metallicity, electron density, gas temperature, kinetic energy, kinematic state) as well as measure the intensity of the ionising radiation front, and the nature of feedback (mechanical, shocks, photoionisation). Theoretical arguments as well as observations of these regions in the Local Group galaxies show that these systems have sizes significantly below 100 pc, suggesting that at typical observing conditions from the ground we can only resolve them within distances of 5 Mpc \cite{McLeod2019,McLeod2021}. Studying HII regions, SNRs, PNe in the Local Volume up to 5 Mpc distant galaxies is thus a benchmark for probing HII regions physics and conditions in the Local Universe and beyond \citep{dellabruna2022}. Within this distance, it is possible to cover main sequence galaxies from $10^8$ to $10^{11}$ Msun, representative of the star formation spectrum of the local universe in terms of galactic environments and metallicity ranges (from a few percent to supersolar). An impact in the studies of the baryonic cycle from tens of parsecs out to galactic scales is possible with an IFS having a large FoV, high sensitivity, moderate spectral resolution and covering blue wavelength regions. Current instrumentation is missing a part of the optical spectrum ($<4800$ \AA), lacking key diagnostic lines necessary to probe gas physical conditions, and spectral resolution for kinematic information. BlueMUSE coverage will allow us to observe at 10 pc scales (or better) emission lines sensitive to temperature and electron density enabling spatial mapping of abundances, and correlation with the ionising source. Moreover, resolved kinematics ($\sim$10 km/s with a spectral resolution of 35 km/s) will enable us to map the expansion of HII regions and estimate coupling efficiency between stellar feedback and ionised gas, as well as molecular gas, routinely sampled with high-spectral resolution in local galaxies  \citep{leroy2021}.

\noindent
\textbf{Resolved stars in GCs and integrated light of star clusters}

3D spectroscopy, assisted with high spatial resolution imaging, allows detailed spectroscopic investigation for tens of thousands of individual stars in globular clusters and heavily crowded cluster centers. Recent works using MUSE increased the spectroscopic samples by two orders of magnitude, enabling detailed studies of the kinematics of the clusters (e.g. \cite{Kamann2018,Kamann2018b}), binary searches \cite{Giesers2018}, and the measurement of stellar parameters \cite{Husser2016}. It is now well established that globular clusters have at least two distinct populations with differences in light elements like sodium or oxygen, and possibly helium (see \cite{Bastian2018}, for a review). 

BlueMUSE will bring a significant improvement in understanding the formation and evolution of star clusters. Low-mass stellar spectra have significantly more spectral lines in the visual-near-UV spectral range covered by BlueMUSE. A spectral coverage down to 350 nm will enable the separation of Nitrogen reach and poor populations with low-resolution spectroscopy (e.g. \cite{Hollyhead2017}). Furthermore, BlueMUSE will allow detailed studies of the blue stars in GCs – such as horizontal branch and potentially blue hook stars, extremely low mass white dwarfs, or interacting binaries. As these objects probe the binary evolution in dense stellar populations their study is important for understanding the overall dynamical evolution of the cluster. 

Integrated light of star clusters (beyond the Local Group) will contain important spectral features (Wolf-rayet features, absorption lines similar to those of massive stars described above, Balmer break and absorption lines) which will enable us to derive ages and metallicities of these systems. High spectral resolution kinematics offered by BlueMUSE can help determining the dynamical mass. 

\noindent
\textbf{Comets}

Comets are pristine relics of the protoplanetary disk where the planets formed and evolved. They hold vital clues about the early solar nebula within their nuclei. When a comet approaches the Sun, the ices in its nucleus start to sublimate, to form an atmosphere around the nucleus called the coma. Because the coma of comets is extended, large field of view IFUs are invaluable to study them. The coma's extended nature makes large field of view Integral Field Units (IFUs) crucial for studying the physical and chemical properties of cometary comae. These IFUs allow us to map the morphology of gaseous species in the coma to study the underlying processes that shape it, and set constraints on the intrinsic properties and composition of the nucleus, and the processes governing the release mechanisms of radicals in the coma. Large field of view IFUs have also proven extremely useful to detect faint activity levels in comets, helping us understand how their activity evolves with varying distances from the Sun. A number of radicals are observable in the blue part of the optical range: CN (388nm), C2 (517nm), C3 (405nm), NH2 (500-700 nm), N2+(391 nm), CO+ (400 and 425 nm) (and NH around 335 nm). Observing these radicals with BlueMUSE allows us to gain in-depth insight into the production mechanisms of species in the coma of comets, as well as their activity. 
In addition to comets from our own solar system, BlueMUSE will offer unique opportunities for the observation of interstellar objects (ISOs), which formed in other planetary systems and pass through our own solar system. Observations of these interstellar comets with a sensitive IFU will allow us to characterise their composition and potentially discover new emission lines from elements not detected in solar system comets. Compared to what is known about the composition of solar system comets, studying ISOs will bring invaluable information on the formation conditions of planetesimals in other planetary systems. 
Observations with BlueMUSE will nicely complement those with CUBES, the upcoming near-UV spectrograph for the VLT, that will cover bluer wavelengths (300-400nm) not accessible with BlueMUSE.

\subsubsection{Nearby Galaxies}
\textbf{Key Science Case: ISM and HII regions in extreme starbursts}

Local low-metallicity (extreme) starburst galaxies offer an opportunity to study galaxy formation under conditions similar to those of the first galaxies. Such galaxies contain rare massive stars, are likely to host superluminous supernovae (hypernovae), have “porous” ISM which enhances the escape of Ly$\alpha$ and LyC radiation, and they enable studies of star formation and its feedback effects, including outflows and the enrichment of the IGM. In these respects, these galaxies are close analogues of the faint galaxies in high-z surveys, that dominate the SF budget. The goal of this science case is to study intense starbursts in great detail, mapping the physical properties at a sub-kpc level, while encompassing the full extent of the target(s).

\noindent
\textbf{Low Surface Brightness (LSB) galaxies}

Given the faintness of the LSB galaxies it is not surprising that, while they might comprise up to 50\% of local galaxies, their general properties, nature and origin are still poorly known. We still need to understand what is the shape of their luminosity function and how it combines with the general luminosity function of z=0 galaxies. LSB galaxies are sources of star formation in a low-density regime, which might be reflected in the initial mass function. The stars or emission-line gas in the LSB galaxies are likely to move almost exclusively under the gravitational influence of the dark matter, and therefore these galaxies are excellent probes of dark matter models, as well as modified gravity theories. Recent discovery of baryon rich LSB, with baryon fractions at the level of the cosmic average \cite{2019ApJ...883L..33M}, indicate a possibility for the existence of alternative dark matter flavours, such as the self-interacting dark matter or the fuzzy dark matter \cite{2024arXiv240406537M}. Giant LSB galaxies, such as Malin 1, with big ($\sim$50kpc), but faint ($<$25 mag/arcsec2) stellar disks are relatively rare, with about 40 known such \cite{Saburova2023}. Numerical simulations predict that 6\% of galaxies in the mass range 10.2$<$log(M/M$_\odot$)$<$11.6 are galaxies of such kind, with $\sim$200 cases found in IllutstrisTNG 100 suite \cite{Zhu2018}. Based on the known cases an approximate number density is giant LSBs is $4\times10^{-5}$ Mpc$^{-3}$ within $z<0.1$, projecting into several thousand candidate galaxies observable from Paranal. This science case focuses on the census of the star formation and dynamical properties of LSB, as probes of galaxy evolution.

\noindent
\textbf{Environmental effects in local galaxy clusters}

The environment plays a major role in shaping galaxies and determining their subsequent evolution. The range of possible physical processes that operate is large, but we are missing a clear picture of which process is dominating a certain type of transformation. Moreover, while it is clear that the processes can be, broadly speaking, either linked to the gravitational perturbations or to hydrodynamical interactions (for example of the cold ISM and warm IGM), their relevance with respect to the environmental density, galaxy (stellar) mass and the interaction epoch (local or inter-z universe) remains a major challenge to our understanding. 

\noindent
\textbf{High order kinematics of cold stellar disks and dwarf galaxies}

Detailed dynamical studies of stars in present-day galaxies provide a fossil record of their individual assembly history. In particular the high-order kinematic moments, i.e.\,the skewness and kurtosis of the line-of-sight velocity distribution (LOSVD), reveal complex stellar orbital structures that go undetected when measuring the velocity and velocity dispersion alone. High-order kinematic measurement will be crucial for revealing the origin of the unusual low ratios of $V/\sigma$ in low-mass spirals and spheroidals ($\log \mathrm{M}^*/\mathrm{M}_\odot <9.5$) that break away from fundamental galaxy scaling relations (e.g., Faber-Jackson, M$^*$-S$_{0.5}$). Similarly, Galactic archaeology surveys are currently revealing that the Milky Way’s disk is shaped through minor collisions with satellite galaxies leaving measurable stellar kinematic signatures. To place our Milky Way in a cosmological context, high-precision measurements of the stellar orbits in galactic disks are required to understand how minor mergers impact the formation and evolution of galaxy structure.

\noindent
\textbf{Elliptical galaxy stellar populations from integrated light spectroscopy in the far blue}

The far-blue spectrum of elliptical galaxies (and other passive populations) gives access to a number of interesting features, including the CN band at 388 nm and the NH band at 336 nm. The latter is probably the “cleanest” nitrogen abundance indicator available in old galaxies, avoiding the complicated molecular interplay between C, N, and O which plagues the interpretation of the more-easily observed CN bands. Previous work on galaxy centers indicated a very weak dependence of NH on velocity dispersion, in contrast to results from CN bands at longer wavelengths; limited spatially-resolved work with X-SHOOTER likewise suggests weaker radial gradients than derived from redder features. Starting with existing archival observations of nearby early-type galaxies (i.e., MUSE data), BlueMUSE would add and improve the abundance-ratio information on critical elements such as C, N and O \cite{Conroy2014}.
Beyond the abundance ratios, the far-blue is sensitive to the presence of any very low-level young populations, and to the low-metallicity stellar population. Composite populations may be responsible for the apparent discrepancy with the redder bands, which measure the dominant high-metallicity population. The far blue also probes exotic contributions to the integrated light, including blue and extreme horizontal branch stars.

\noindent
\textbf{Planetary nebulae for stellar populations and kinematics of elliptical galaxies, as well as distance measurement}

Planetary nebulae can be used as resolved point tracers of kinematics out to large radius from their parent galaxy, or into intracluster regions. In addition, the relative number of PNe per stellar mass of the underlying population can be an interesting probe for post-RGB evolution, especially at the high metallicities found in the cores of ellipticals. The challenge (especially in the latter case) is to find PNe against the high continuum from the normal stellar population. IFS observations are well suited for such work by their ability to “zoom in” to the individual emission-line, over a large area. Nevertheless, the galaxy cores remain challenging. Contrast is key to detecting PNe here: either spatial compactness (high spatial resolution) or the narrowness of the lines (high spectral resolution) can be exploited to reduce the background noise.
The Planetary Nebula Luminosity Function (PNLF) is also an established distance measurement technique \cite{Ciardullo2012} that, however, has not become an independent tool on the distance ladder towards the determination of the Hubble constant H$_0$ because the limiting distance, based on narrow-band filter techniques, did not extend so far beyond $\sim$15 Mpc. A proof-of-principle with MUSE \cite{Roth2021} allowed to reach galaxies out to 30 Mpc with 0.04 mag precision. BlueMUSE will enhance that capability significantly due to high efficiency in the blue spectral region that is important for the detection of [OIII] 5007 \AA\ and [OIII] 4959 \AA\ , as well as H$\beta$, all of which are important for this technique. Depending on the expense of observing time, distances as far as 50 Mpc come into reach. [OIII] 4959 \AA\ is specifically needed to differentiate background LAEs. 

\noindent
\textbf{Soft X-ray sources and radiative shocks in metal-poor galaxies}

Metal-poor galaxies are known to harbor relatively harder radiation fields compared to their metal-rich counterparts but the culprit sources are still heavily debated. Such a hard radiation field has been found in the nearby (e.g., \cite{Guseva2000,Schaerer2019,Plat2019} but also in the distant (e.g., \cite{Stark2016}) Universe. Current stellar evolution and atmosphere synthesis models predict too soft radiation fields, and, as a result, the intensity of spectral lines such as HeII 4686\AA\ in the optical or [OIV] 25.9$\mu$m in the infrared (requiring photon energies in the range 54-77eV) is so far unexplained (e.g., \cite{Stanway2019, Nanayakkara2019}. This leads to unacceptable uncertainties on the radiation field of most galaxies in the Early Universe which are metal-poor and star-forming. 

The small statistics of high-ionization lines in nearby metal-poor galaxies prevents unambiguous identifications of the ionization mechanisms at work. To identify the ionization mechanisms, we need not only interstellar medium models adapted to low-metallicity conditions but also better observational constraints. In particular, we need to estimate the precise location of HeII emission and total He+-ionizing flux within the most nearby metal-poor galaxies, and more generally to increase the sample of low-z galaxies with HeII and [NeV] detections in the optical.
Extending the wavelength range down to $\sim$3400\AA\ currently provides the best avenue to detect the most constraining high-ionization lines. In the most nearby galaxies, one can also use such capabilities to map ionization nebulae around specific soft X-ray sources such as WR and ULX in the fashion of, e.g., \cite{Gurpide2022,Gruyters2012}.

\subsubsection{Distant Universe and Deep Fields}

\noindent
\textbf{Key science case: Gas flows around galaxies in deep fields}

Despite the impressive success of recent large-scale cosmological simulations at reproducing the bulk of galaxy properties across cosmic time, the fundamental questions about how galaxies acquire gas from the intergalactic medium, and how they regulate their growth through galactic winds or other preemptive processes, are mostly unconstrained from observations or theory. Predictions of outflow rates from different state-of-the-art simulations differ by orders of magnitudes. From this, it is also clear that observing the Circum-Galactic Medium (CGM) and constraining the flows of gas that traverse it would be a radically new constraint on galaxy formation. MUSE has spectacularly demonstrated the ability of panoramic IFS to observe the CGM through its emission in the Lyman-$\alpha$ line of hydrogen (e.g., \cite{Wisotzki2016}, \cite{Guo2023}). Observations with BlueMUSE in the redshift range $2 < z < 3$ will have a key advantage for constraining galaxy formation because they cover the peak of cosmic star formation. Among the factors that drive this turn-over in the cosmic SFR, we may expect a change in the form of accretion flows and their interactions with galactic winds, which marks the beginning of the transition from the early Universe (where gas flows cold and collimated onto galaxies), and the late Universe (where star formation is sustained by cooling-regulated accretion from hot coronae). Observing the evolution of the CGM through this epoch will be a key in discriminating between the various competing theoretical models.

\noindent
\textbf{Gravitational Lensing by massive clusters}

Massive galaxy clusters locally curve Space-Time and thus stretch and magnify the light of background galaxies. Within the central square arcminute, distant sources are generally multiply imaged by the lensing effect and the magnification of these images is typically larger than a factor of a few but can reach factors of hundreds for gravitational arcs straddling the critical lines. Thanks to the lensing magnification we can study lower luminosity/mass galaxies that would otherwise be undetectable, and spatially resolve the internal structure of distant galaxies at sub-kpc scales.
With its wide FoV and blue wavelength range, BlueMUSE will have a unique capability to probe a large number of such magnified sources at the very faint-end of the Lyman-alpha luminosity function (LF) at $z \sim 2-3$, thus testing the presence of a turnover in the LF around cosmic noon (e.g. \cite{Alavi2016}). Such faint star-forming sources are expected to be analogs of the sources of reionisation.
Based on previous lensing work at higher redshift with the MUSE instrument (e.g. \cite{Richard2021}) we can also extrapolate the redshift range $z \sim 2-4$ to be also a peak in the number of multiple images and highly magnified sources. As a consequence, the number of giant arcs (stretched typically by more than 10”) lensed by massive clusters is expected to also be very significant (e.g. \cite{Hennawi2008,Bayliss2011,Johnson2017} in this redshift range. This makes BlueMUSE the perfect instrument to probe resolved gas properties illuminated by extended bright sources: (i) resolved gas absorption and emission within the distant giant arc (e.g. \cite{Patricio2016,Claeyssens2019}), (ii) gas absorption in the CGM and IGM along the line of sight (e.g. \cite{Lopez2020}).
In addition, the combination of BlueMUSE and MUSE observations of the same cluster fields will provide the largest number of multiple images with precise spectroscopic redshifts, allowing for a precise mapping of the total mass distribution in the cluster cores including the unknown dark matter (e.g., \cite{Lagattuta2017}), with potential further application to put constraints on the properties of dark matter cross-section (e.g., \cite{Harvey2017}) or cosmological parameters (e.g., \cite{Jullo2010}). 

\noindent
\textbf{Forming Groups and clusters}

BlueMUSE will open the Ly-$\alpha$ redshift window 1.87$<z<$4 to study cold gas in early over-dense environments (predominantly expected to be groups and clusters) over 500-700 kpc scales. Open questions that can be addressed in this crucial redshift range include: the origin of the gas, the powering sources of Ly-$\alpha$ emission and other emission lines such as CIV, HeII, and MgII, the role in the evolution of galaxies, and the energy between phases, particularly with the hot low density medium. Crucial to this research is the detection of line emission down to low surface brightness levels to probe filaments of the cosmic web as part of the overall large-scale structure. 

BlueMUSE observations will provide a unique way to constrain cold stream accretion predicted by theory, but also study feedback from AGNs (also in the form of outflows), gas recycling, and the physics of multiphase circumgalactic medium. This science will likely require a tiered approach: from global statistics with moderately deep observations to in-depth studies of a few carefully selected objects to mapping larger areas of the sky to investigate connections between structures within the cosmic web. Selection of appropriate targets cluster/group samples will be critical: Euclid/WFIRST, JWST, new Sunayev-Zeldovich probes, Xrays (Athena will play a crucial role that we need to prepare, but will come later) but also large spectroscopic surveys (e.g., MOONS, 4MOST, WEAVE, etc).
Observations with BlueMUSE will allow for crucial tests of models of structure growth and numerical simulations as well as tests of multiphase gas physics in the halos of galaxies, groups, and overdensities.

\noindent
\textbf{Lyman-continuum leakers}

We seek to measure the ionizing output from galaxies at mid-redshifts to (a.) determine the metagalactic ionizing background, and (b.) determine the escape fraction of ionizing photons for application in the reionization epoch. The galaxy luminosity function is steep and most of the ionizing photon budget ($\dot{n}_{\rm ion}$) is produced by faint galaxies. Measuring flux at $\lambda <91.2$ nm rest-frame cannot be done at $z>4$ because of the intervening intergalactic medium, nor at $z<2.7$ because of the atmospheric cutoff (HST studies are excellent but biased by photometric pre-selection and hard to apply to high redshifts).
BlueMUSE will (1) collect a statistical sample of Lyman Continuum (LyC) Emitters to investigate the analogues of the sources of cosmic reionisation at 2.7 < z <4 (2) test observational diagnostics for LyC leakage based on the Lyman-$\alpha$ properties (3) constrain the physical properties of typical LyC Emitters. 

\subsection{Uniqueness and synergies of BlueMUSE} 
 
The technical capabilities of the BlueMUSE instrument will make a strong complement to the existing and future instruments at the VLT. As a panoramic IFS, it will naturally complement imagers, multi-object and long-slit spectrographs. Particular synergies are envisioned with MUSE (similar FoV, but with lower spectral resolution and optimised to the red), CUBES (high resolution long-slit spectrographs optimised for the UV) and MAVIS (high spatial resolution imager and small FoV IFS). Similar synergies can be expected between BlueMUSE and Extremely Large Telescope instruments (there is no IFS efficiently covering the wavelength range below 500 nm on the ELT), with BlueMUSE bringing the blue wavelength coverage and large FoV IFS.

The nearest instrument in terms of performance is the Keck IFU KCWI instrument, but BlueMUSE with 20 $\times$ larger field of view at the same spectral resolution, better throughput and stability and an overall increased efficiency, will be one to two orders of magnitude more efficient.

At a time when the focus of most of the new large facilities (ELT, JWST) will be on the infrared, BlueMUSE will be a unique facility outperforming any ELT instrument in the Blue/UV. It will have a strong synergy with ELT, JWST as well as ALMA, SKA, Euclid and Athena.

\subsection{Top Level Requirements} 

Each of the science cases presented before are used to derive some of the main performance requirements in Table \ref{tab:tlr}: wavelength range, spectral resolution, field of view and spatial sampling. For each of them, the global instrument TLR has been selected to fulfill, at least partially, all of these science cases. 

The common wavelength coverage of the BlueMUSE instrument (defined as the minimum wavelength range covered by the detector at any spatial position in the Field of View) shall fully contain the wavelength domain 350 nm $<\lambda<$ 580 nm.  However we define as ‘Extended wavelength range’ the additional wavelengths acquired on the detector but only partially covered by the Field of View (approximately a third of the FoV from the current design). The limits in wavelength of the extended range are expected to be larger, $\sim330$ nm $<\lambda<600$ nm and are set as a goal in order to recover additional parameter space for this TLR.

\begin{table}[ht!]

\begin{tabular}{lclll}
Science Case & \begin{tabular}{l}Wavelength\\ range (nm)\end{tabular} & \begin{tabular}{l}Average\\spectral\\resolution\end{tabular}& FoV (arcmin$^2$) & \begin{tabular}{l}Spatial\\sampling (")\end{tabular} \\\hline
\textcolor{red}{\it SWG1: Local Volume}\\\hline
\rowcolor[rgb]{0.8,0.8,1.0}\begin{tabular}{l}Wolf-Rayet (WR) and OB star\\ evolution\end{tabular} & 350-580 & $>=3500$ & $>1$, goal 2 & $0.2<s<0.3$\\\hline

\begin{tabular}{l}Resolved HII regions, SNRs,\\ and PNe\end{tabular} & 350-580 & \begin{tabular}{l}$>=3500$, $>2800$\\ at 370 nm\end{tabular} & $>1$, goal 2 & $0.2<s<0.3$\\\hline

\begin{tabular}{l}Resolved stars in GCs and \\ integrated light of star clusters \end{tabular} & \begin{tabular}{c}350-580\\(goal 336)\end{tabular}& $>=3500$ & $>1$, goal 2 & $0.2<s<0.3$\\\hline

\begin{tabular}{l}Comets\end{tabular} & \cellcolor[rgb]{1,1,0.7}\begin{tabular}{c}335-520\end{tabular}& $>=1000$ & $>1$ & $0.2<s<0.3$\\\hline

\textcolor{red}{\it SWG2: Nearby Galaxies}\\\hline

\rowcolor[rgb]{0.8,0.8,1.0}\begin{tabular}{l}ISM and HII regions in extreme \\ starbursts\end{tabular} & 350-580 & \begin{tabular}{l}$>=3500$, $>2800$\\ at 370 nm\end{tabular}  & $>1$ & $0.2<s<0.3$\\\hline

\begin{tabular}{l}Low Surface Brightness (LSB) \\galaxies\end{tabular} & 350-580 & $\sim3000$ & $>1$ & $0.2<s<0.3$\\\hline

\begin{tabular}{l}Environmental effects in local \\galaxy clusters\end{tabular} & 350-580 & $\sim3500$ & $>1$ & $0.2<s<0.3$\\\hline

\begin{tabular}{l}High order kinematics of cold  \\stellar disks and dwarf galaxies\end{tabular} & 375-575 & $\sim3500$ & $>1$ & $0.2<s<0.3$\\\hline

\begin{tabular}{l}Elliptical galaxy stellar \\populations from integrated light \\spectroscopy in the far blue\end{tabular} & \cellcolor[rgb]{1,1,0.7}336-480 & $\sim$3500 & $>1$ & Not critical\\\hline

\begin{tabular}{l}Planetary nebulae for stellar \\populations and kinematics of \\ elliptical galaxies + distances\end{tabular} & \begin{tabular}{l}Including\\500.7 nm\end{tabular}& $3500$ at 500 nm & $>1$ & $0.2<s<0.3$\\\hline

\begin{tabular}{l}Soft X-ray sources and radiative\\shocks in metal-poor galaxies\end{tabular} & \cellcolor[rgb]{1,1,0.7}\begin{tabular}{l}340-580\\(including \\342.6 nm)\end{tabular}& $\sim3000$ & $>1$ & $0.2<s<0.3$\\\hline

\textcolor{red}{\it SWG3: Distant Universe}\\\hline

\rowcolor[rgb]{0.8,0.8,1.0} Gas flows around galaxies & 350-580 & \begin{tabular}{l}$2800<R<$
\\3700\end{tabular}& $>1$, goal 2 & $0.2<s<0.3$\\\hline

\begin{tabular}{l}Gravitational Lensing by \\massive clusters\end{tabular} & 350-580 & \begin{tabular}{l}$2500<R<$
\\4500\end{tabular}& $>1$ & $0.2<s<0.3$\\\hline

Forming Groups and clusters  & 350-580 & & $>1$ & $>=0.3$\\\hline

Lyman-continuum leakers  & \cellcolor[rgb]{1,1,0.7} 350-600 & \cellcolor[rgb]{1,1,0.7} $<3500$& $>1$, goal 2 & $>=0.3$\\\hline

{\bf Derived TLR} & {\bf 350-580 nm} & \begin{tabular}{l}\bf Min R.$>2600$\\{\bf Avg R.$>3500$}\end{tabular}&{\bf$>$1 arcmin$^2$}&{\bf 0.2"$<s<$0.3"}\\

\end{tabular}

\caption{\label{tab:tlr}Summary of the main instrument performance requirements from the science cases, as well as the derived TLR at the bottom. One key science case in each of the 3 science categories is highlighted in blue, and their requirements shall be entirely fulfilled by the instrument TLR. Pale yellow cells mark requirements only partially covered by the derived TLR.}
\end{table}

In addition to the main TLR presented in Table~\ref{tab:tlr}, BlueMUSE has to fulfill the generic requirements specified for all VLT instruments (telescope interface, software architecture, ...). We also include TLRs aiming at maximising the overall science return (not science case specific), such as:

\begin{itemize}
    \item{{\bf End-to-end throughput}: the minimum end-to-end throughput (including all transmission losses due to the atmosphere at the zenith, telescope  and the BlueMUSE instrument) shall be higher than 15\% across the common wavelength range. The average end-to-end throughput within the common wavelength range shall be higher than 26\%.}
    \item{{\bf Image quality}: the BlueMUSE instrument point spread function (including all optical aberrations and contributions from derotator wobble and differential atmospheric dispersion during an exposure, but excluding telescope) shall have a FWHM lower than 0.42" in the common wavelength range.}
    \item{{\bf Stability}: in operational conditions, the BlueMUSE instrument shall be stable in spatial and spectral position to allow for the calibration of sky position and wavelength within 0.1 pixel (including the effect of derotator wobble), illumination within 10\%, based on calibrations taken during the day or twilight.}
    \item{{
    \bf Spectral sampling}: The pixel sampling of the BlueMUSE instrument detectors shall sample the line spread function (LSF) FWHM along the spectral direction by at least 2 pixels (goal: 2.5 pixels) at any wavelength of the common range.}
    \item{{\bf Operations efficiency}: The BlueMUSE instrument shall have a ratio between the open shutter time and the execution time of the observations (including telescope and instrument overheads excluding all unexpected technical downtime) higher than 70\%.}
\end{itemize}

\section{System architecture}
\label{sec:architecture}

The overall instrument architecture of BlueMUSE follows the key points of MUSE, including a two-steps approach to the image slicing which allows to cover the large number of slices required to sample the field of view. To achieve its top level function and cover its lifecycle, BlueMUSE is split into several subsystems. We present here an overview of the main subsystems in order of appearance in the optical path from telescope to detector, including a summary of its conceptual design when available, as well as the main challenges. 

\subsection{Calibration Unit}

The Calibration Unit (CU) is an opto-mechanical structure with calibration lamps and associated electronics. Its task is to uniformly illuminate BlueMUSE with a calibration beam that mimics both the VLT f-number and its pupil illumination. It includes functions to switch on and off the lamps and motors to position calibration masks into the FoV. A calibration pick-up mirror is used to direct light into the instrument from either the telescope or the CU. Flat-field lamps shall provide relatively uniform continuum flux over the BlueMUSE wavelength range, while spectral line sources are used for wavelength calibration.

\subsection{Fore-Optics}

The Fore Optics (FO) reshapes the VLT focus image to adapt it to the next subsystem, which splits the image into 16 channels. It rotates the field using a classical derotator, then magnifies it and reshapes it to a 1.33:1 ratio. This anamorphosis is necessary to properly sample each slice onto the detector. 

\begin{figure}[ht!]
\begin{minipage}{9cm}
\includegraphics[width=9cm]{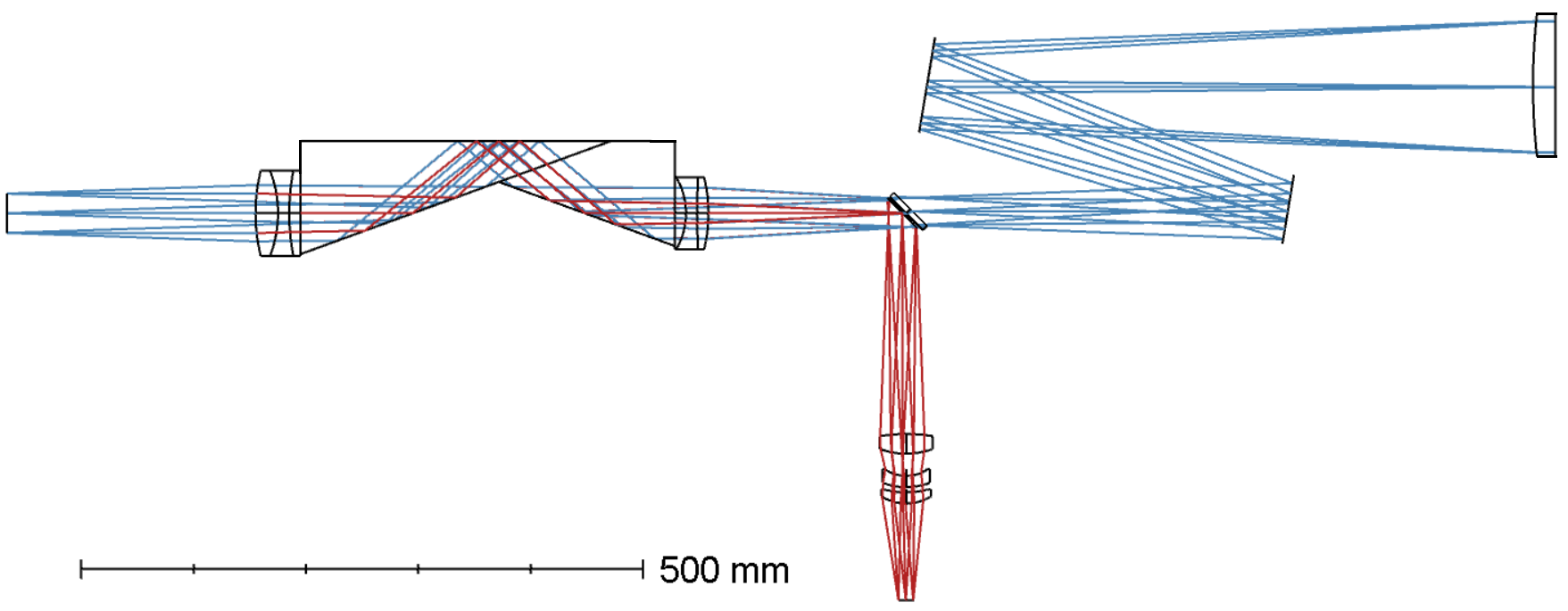}
\end{minipage}
\begin{minipage}{8cm}
    \caption
    { \label{fig:fo} 
 Schematic view of the FO optical design: the science light path is presented in blue, while the redder wavelengths feeding the secondary guiding / monitoring camera are presented in red.}
\end{minipage}
\end{figure}

Compared to the MUSE Fore Optics design, the BlueMUSE FO is optimised for blue/UV wavelengths, and the anamorphosis factor is smaller. There is no scale changer for narrow-field mode, no Atmospheric Dispersion Correction (ADC), and no filter wheel needed. A schematic view of light rays in the FO is presented in Fig.~\ref{fig:fo}.

In addition to the main science path presented in this figure as blue light rays, a $\sim$600 nm shortpass dichroic separates the redder wavelengths light rays and they are reimaged with a camera onto a technical CCD. The simultaneous acquisition of images can be used either as a monitoring system or secondary guiding for BlueMUSE data. Unlike the MUSE slow-guiding system, this approach allows to cover the entire science field of view and have much less optical distortions. It is also smaller/lighter and more cost-effective.

\subsection{Splitting and Relay Optics}
The Splitting and Relay Optics (SRO) system splits the BlueMUSE FoV into 16 channels and redirects the light of each channel towards the entrance of an IFU. It is composed of a field-splitter and a field separator, which separate the field into 16 horizontal beams. These sub-fields feed 16 relay optics, which correct for the variations in path length from one channel to another.

The design of the SRO system has to be adapted compared to MUSE according to the number of channels and f-number. It is foreseen to use micromechanical systems in the relays in order to improve IFU alignment during AIT, maintenance and day-time calibrations.

\subsection{Structure and Enclosure}

The main structure of BlueMUSE is a mechanical system that provides monolithic structural supports to all opto-mechanical sub-systems. It is the single mechanical system interfacing the Nasmyth platform. It allows for the overall instrument stability with time and temperature.

\begin{figure}[ht!]
\begin{minipage}{9cm}
\includegraphics[width=9cm]{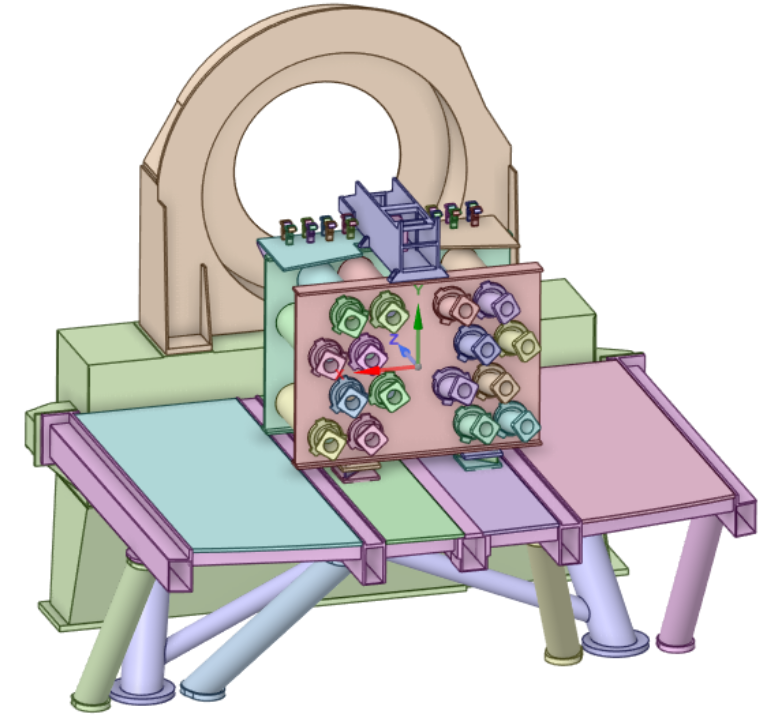}
\end{minipage}
\begin{minipage}{8cm}
    \caption
    { \label{fig:structure} 
 General idea of the mechanical structure of BlueMUSE on the VLT Nasmyth platform, including the 16 IFUs standing out at the back as well as the fore-optics in blue at the top.}
\end{minipage}
\end{figure}

Overall, the structure ought to be similar to the MUSE main structure:  a general idea of its appearance is shown in Fig.~\ref{fig:structure}. In addition, based on the experience obtained with the MUSE instrument \cite{spie_2024cai}, we are studying for BlueMUSE an enclosure which will buffer small temperature fluctuations typically occurring during nighttime observations, by minimising the thermal drifts that can affect optical performance. The proposed system might incorporate active thermal compensation technologies that dynamically adjust to external temperature changes.

This enclosure would also serve as a protective barrier against environmental factors such as dust and moisture, which can degrade the performance of optical components over time. By controlling the internal environment, BlueMUSE aims to operate within a narrower range of temperature variations, thus reducing the frequency and complexity of recalibrations needed to maintain optimal performance.

\subsection{Integral Field Unit}

 Each of the 16 Integral Field Units (IFUs) (assigned to a given channel) is formed by the combination of an image slicer, a spectrograph, and a detector, and their overall architecture is quite identical to their MUSE counterparts.  The BlueMUSE spectrographs and detector systems are significantly different to MUSE and their design is detailed in the next subsections.

\subsubsection{Image Slicer}

\begin{figure}[ht!]
\begin{minipage}{9cm}
\includegraphics[width=9cm]{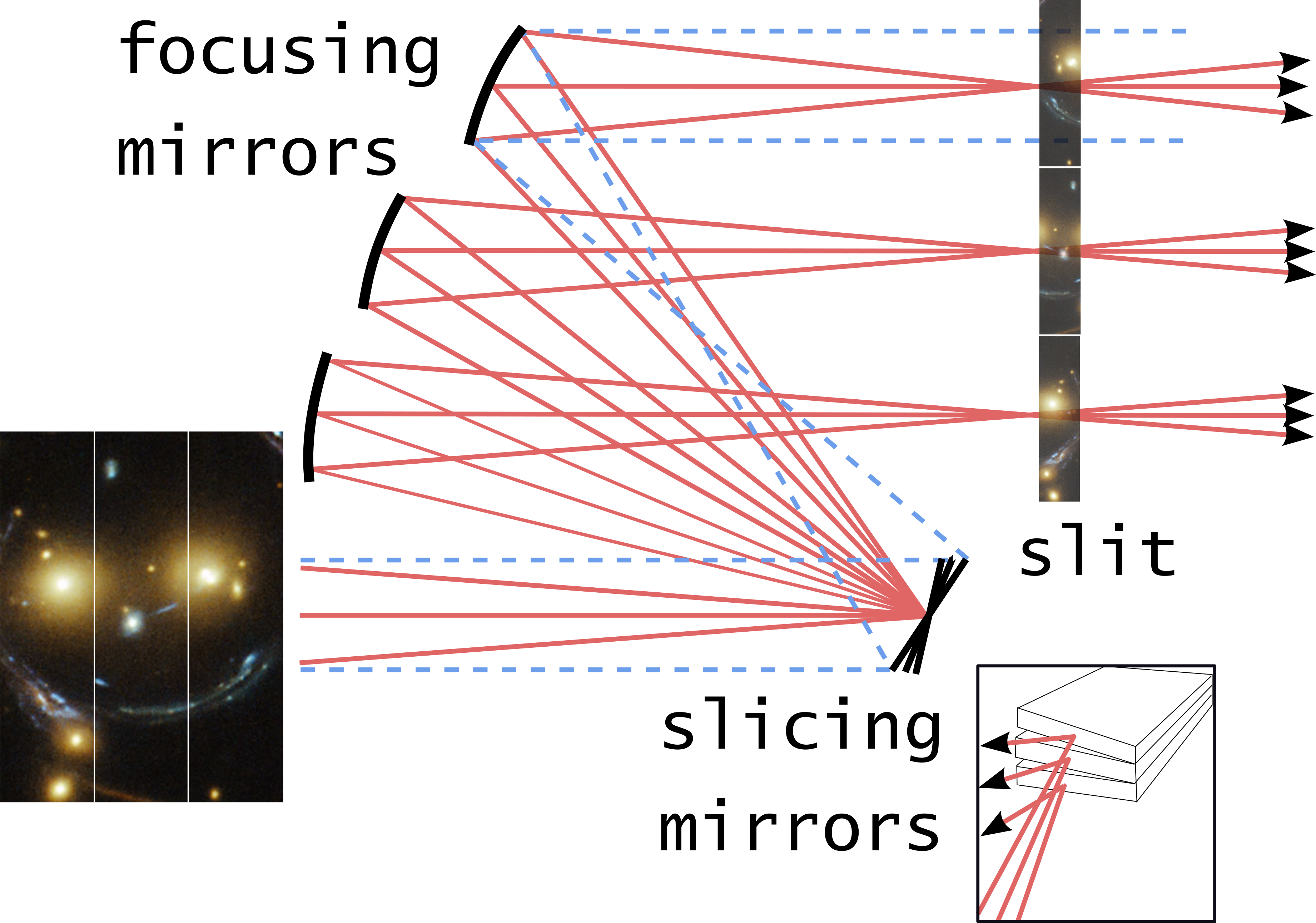}
\end{minipage}
\begin{minipage}{8cm}
\includegraphics[width=8cm]{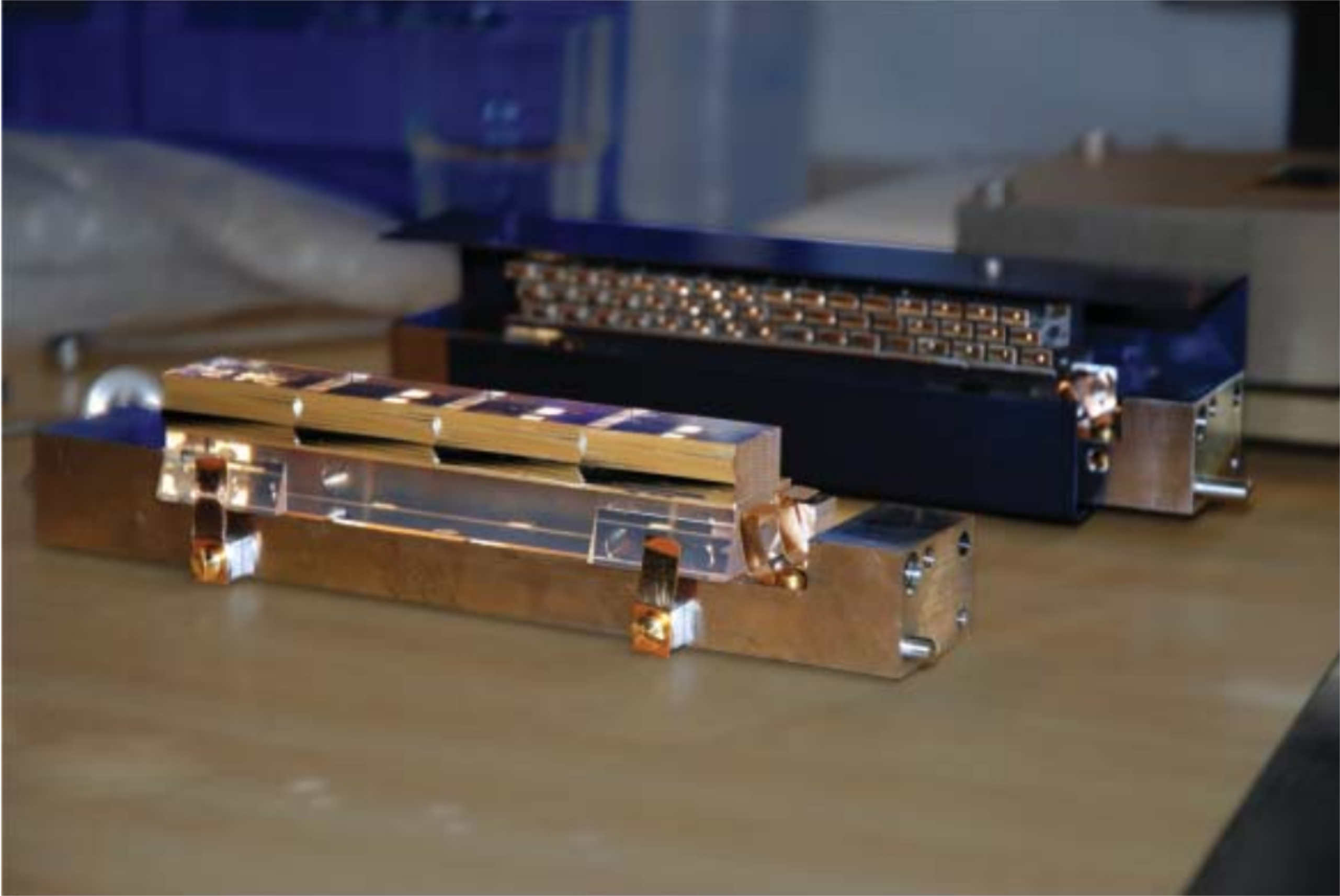}

\end{minipage}
    \caption
    { \label{fig:iss} 
 (Left) Conceptual view of the BlueMUSE image slicer, composed of two arrays of mirrors to produce the pseudo-slit at the entrance of the spectrograph. (Right) Actual picture of a MUSE image slicer showing the two arrays of 48 mirrors. A similar optimisation will be performed for BlueMUSE.}
\end{figure}

\textbf{The image slicer} cuts and rearranges the 2D sub FoV in a 1D pseudo slit of 0.3” width (projected on sky). The main function of the image slicer is to slice the fraction of the FoV coming from the SRO into 48 slits that are rearranged in a long slit at the entrance of the spectrograph. The foreseen slicer design is similar to the MUSE image slicer, which is presented in detail in \cite{Laurent2008} and the concept is summarised in Fig.~\ref{fig:iss}. Compared to MUSE, the choice of an increased slit width will produce rectangular spatial pixels (spaxels) of 0.2" $\times$ 0.3" on the detector. This spatial sampling helps reaching the background-noise limit in the blue under dark conditions, while properly sampling the PSF under the best seeing.

\subsubsection{Spectrograph}

\begin{figure}[ht!]
\begin{minipage}{12cm}
\includegraphics[width=12cm]{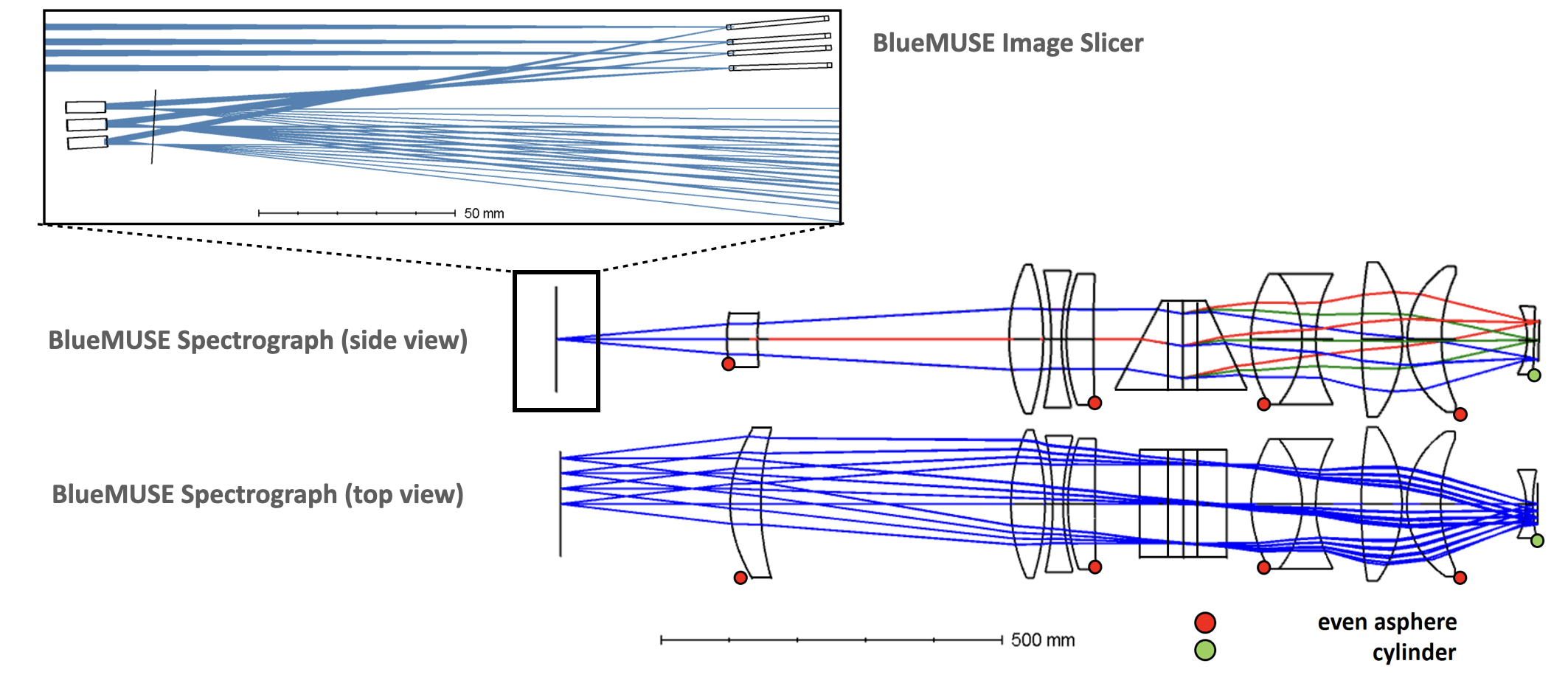}
\end{minipage}
\begin{minipage}{5cm}
    \caption
    { \label{fig:spectrograph} 
 Overall view of the BlueMUSE spectrograph (top and side view) and image slicer optical designs (inset, only 4 slices are represented to improve readability). The color points highlight the non-spherical lenses of the design.}
\end{minipage}
\end{figure}

\textbf{The spectrograph} produces the spectra of the mini slits and images them onto a detector. The optical design of the BlueMUSE spectrograph is based on a similar architecture as the MUSE one, but makes use of a zero-deviation dispersive system, which helps in terms of compactness and installation. More details on the BlueMUSE spectrograph design concept have been presented in \cite{Jeanneau2020}, and an updated version of the optical design is shown in Fig.~\ref{fig:spectrograph}. One of the main challenges of the spectrograph subsystem is the overall throughput of the dispersive element, which makes use of Volume Phase Holographic (VPH) gratings (see also \cite{spie_2024jeanneau}).

\subsubsection{Instrument Detector system}

The Instrument Detector System (IDS) includes the 16 Detector Vessels (DV) mounted on the IFUs, the detector proximity readout electronics, the Vacuum \& Cryogenics System and the overall detector System Control Electronics \& Software. The detectors are flat 4k x 4k, 15 $\mu$m pixel CCDs, operating at 163 K. 
The IDS ensures complete operation of the 16 detectors from detector exposure, readout and exposure FITS file generation.

We are studying a DV solution based on the Cryostat system developed by CEA-IRFU for the 10 DESI spectrographs \cite{DESI2016}. The main difference with regards to the MUSE System is the use of Cryocoolers (Pulse Tube or Stirling, to be defined) instead of Continuous Flow Cryostats and liquid nitrogen. This type of cooling needs only electric power and chiller unit at 10$^\circ$C for operating.

\subsection{Instrument Control Electronics and Software}

The main goal of the instrument control is to provide users with the ability to control and monitor the instrument, including all detectors, moving parts and sensors. The main difference with respect to MUSE is the use of the ELT framework for the control software, to which all VLT2030 instruments have to abide by.

\subsection{Science Software}

The BlueMUSE Science Software is composed of several components:

\begin{itemize}
\item{{\bf The observation preparation tools}: these tools allow the user to estimate instrument performance (e.g. Exposure Time Calculator) and define observations at different stages.}
\item{{\bf Data Reduction Software (DRS)}: the BlueMUSE DRS is derived from the MUSE DRS \cite{Weilbacher2020}, but we are investigating options to include additional improvements such as an improved estimate of the LSF, the propagation of covariance, and proper handling of the BlueMUSE rectangular spaxels. More details on this work are provided in \cite{Weilbacher2022}.}
\item{{\bf Data Analysis Software (DAS)}: during Phase A we are studying the possibility to develop dedicated tools to analyse BlueMUSE (as well as MUSE) datacubes and help astronomers in tasks such as source detection, spectral extraction and source classification.}
\item {{\bf End-To-End simulation software (BlueSi)}: simulating realistic science data cubes as expected from the DRS. 
Integrating all observational and in particular instrumental specifications on cube level, it serves as a testing ground for 
observational strategies and analysis tool chains under the various parameters \cite{spie_2024wendt}.}

\end{itemize}

\section{First view on performance}

\label{sec:perfs}

The first BlueMUSE design presented in Sect.~\ref{sec:architecture} fulfills the main performance TLRs presented in Table~\ref{tab:tlr} in wavelength coverage, spectral resolution, field-of view and spatial sampling (0.2 $\times$ 0.3" spaxels).

We present here a first analysis of other performance TLRs based on the current optical design of the subsystems and the following assumptions:

\begin{itemize}
    \item{Natural seeing: in the absence of adaptive optic system, the image quality will be dominated by the Paranal seeing conditions \cite{Martinez2010}. We assume the following parameters: 0.70" (0.5") for medium (10\% best) seeing conditions defined in the V band, and outer scale $L_0=22$ m.}
    \item{No atmospheric dispersion (AD) correction: we include the contribution of the relative atmospheric dispersion to the image quality budget.}
    \item{Impact of the derotator: there is an expected spatial shift of the image position (wobble) at the level of the derotator subsystem. We use the values of MUSE to estimate the impact of the derotator on image quality during long exposures.}
\end{itemize}

\begin{figure}[ht!]
\begin{minipage}{8cm}
\includegraphics[width=8cm]{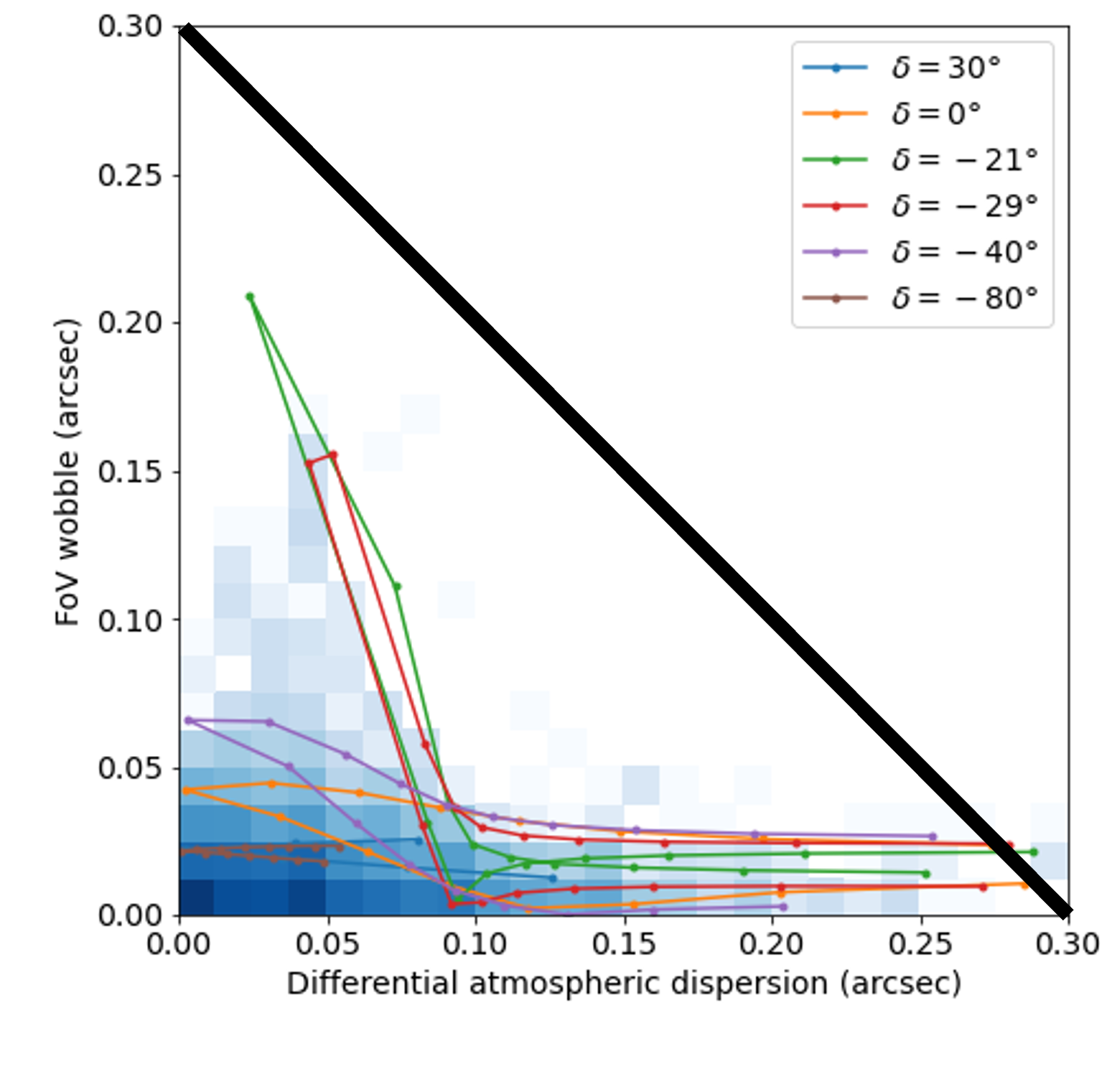}
\end{minipage}
\begin{minipage}{9cm}
    \caption
    { \label{fig:museexp} The blue squares show the distribution of all science exposures taken with the MUSE instrument during 10 years of operations as a function of their estimated
drift coming from derotator wobble (vertical axis) or atmospheric dispersion over the BlueMUSE wavelength range (horizontal axis). 
    The coloured curves show the theoretical estimates for a maximum 30 min. exposure as a function of source declination $\delta$. We can define for BlueMUSE an upper limit for a total 0.3" drift for both effects (black solid line).
 }
\end{minipage}
\end{figure}

As illustrated in Fig.~\ref{fig:museexp}, we do not expect in practice the worst-case contributions of AD and derotator wobble to happen for the same sources, as atmospheric dispersion is stronger at high airmass / low elevation while the wobble is stronger at low airmass / high elevation. Instead we use a total maximum contribution of 0.3" drift for 30 min exposures.

\subsection{End-to-end throughput}
\begin{figure}[ht!]
\begin{minipage}{8cm}
\includegraphics[width=8cm]{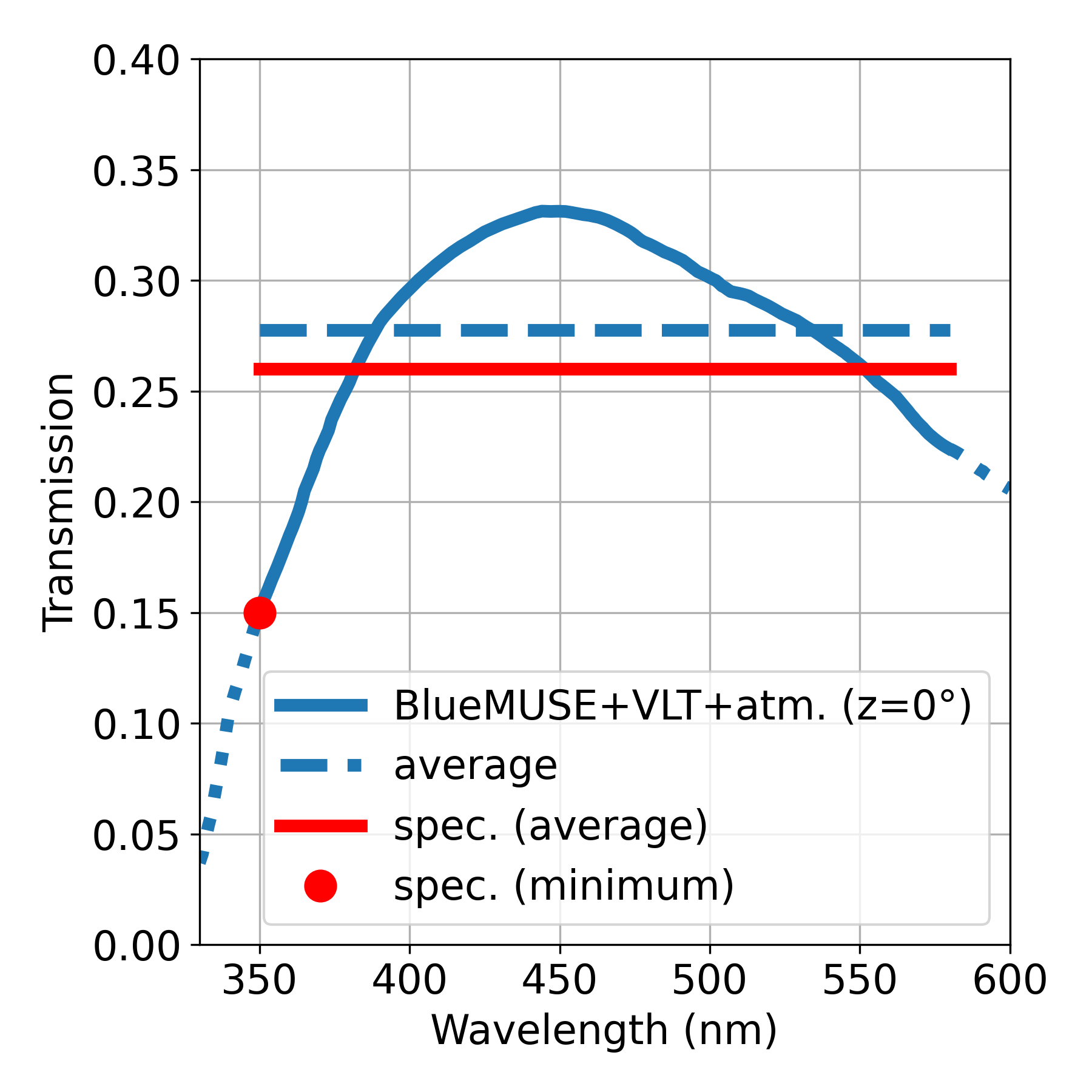}
\end{minipage}
\begin{minipage}{9cm}
    \caption
    { \label{fig:throughputall} 
 The blue curve shows the end-to-end throughput expected from BlueMUSE, including the contribution from the VLT and the atmosphere at zenith, as a function of wavelength. The average value over the common wavelength range is shown as a dashed line. These values can be compared with the minimum and average Top Level Requirements presented in Sect.~\ref{sec:science} and shown in red.}
\end{minipage}
\end{figure}

\begin{figure}[ht!]
\begin{minipage}{9cm}
\includegraphics[width=9cm]{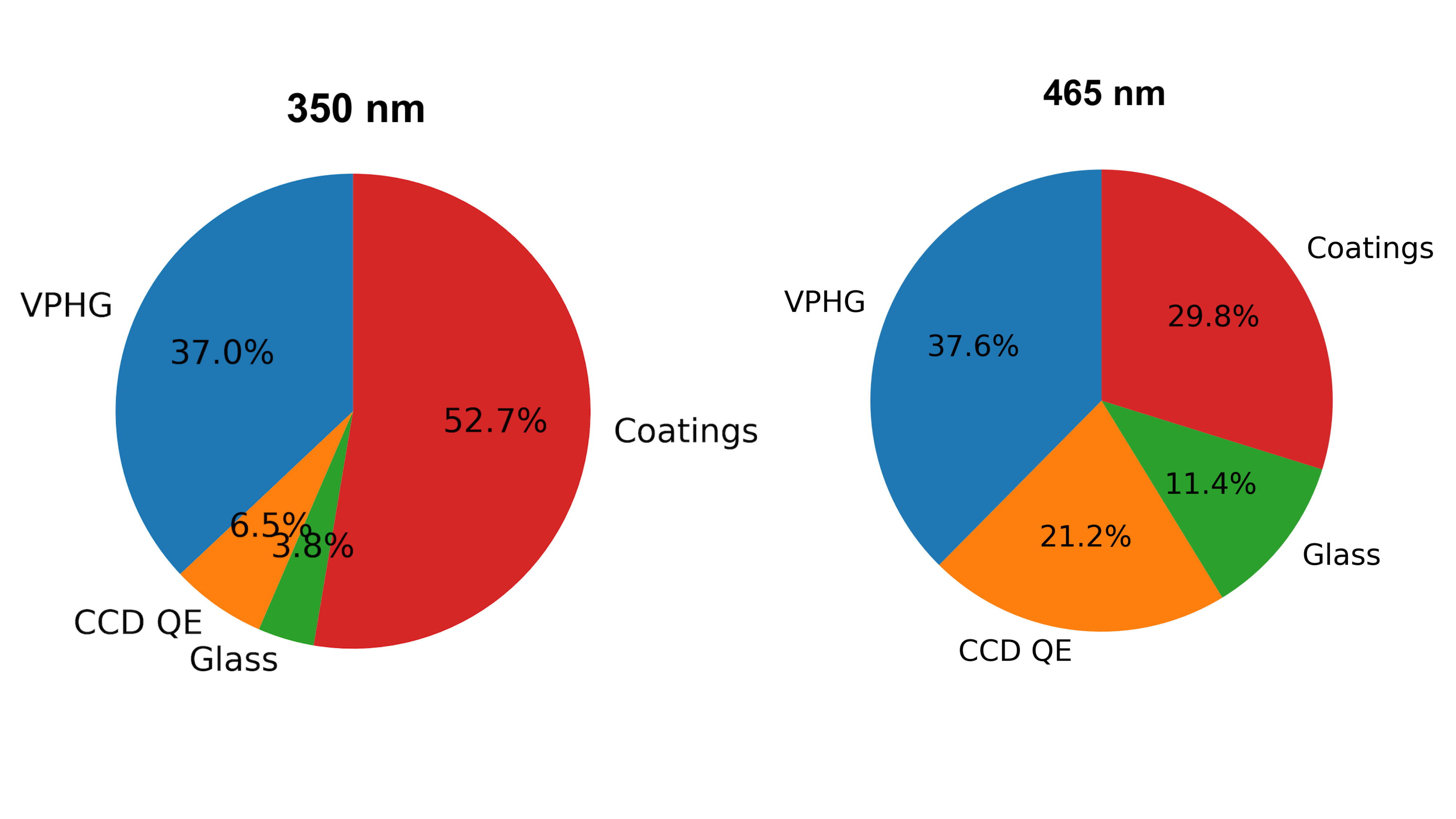}
\end{minipage}
\begin{minipage}{8cm}
    \caption
    { \label{fig:throughput} 
 Relative contributions of different BlueMUSE subsystems to the throughput. We can clearly see the dominance of the VPH grating and detector (CCD quantum efficiency) transmissions at 350 nm.}
\end{minipage}
\end{figure}

The main challenge for BlueMUSE in terms of throughput is the blue end of the wavelength range, with atmospheric transmission dropping significantly, down to 60\% at 350 nm. We use the Paranal extinction curve to account for the additional atmospheric absorption at these wavelengths. We assume a $70\%$ VLT transmission, based on sample measurements from the VLT Coating Unit \cite{1999Msngr..97....4E}. As the wavelength range of BlueMUSE is narrower than the one of MUSE, we can conservatively assume similar coating efficiencies : $>$ 99.3\% for both reflective and anti-reflective dielectric coatings\footnote{Private communication from Optics Balzers}. We assume a metallic coating on the most sensitive mirror array of the image slicer, which yields 93\% for this subsystem. We expect the detectors\footnote{STA4150A, coating A} to perform quite well in the blue (80\% quantum efficiency at 350nm, 95\% on average above 370 nm) and the overall shape of the VPH grating to follow the results from the actual prototype measurements presented by \cite{spie_2024jeanneau}. The overall end-to-end BlueMUSE sensitivity we expect under these assumptions, along with the throughput budget at 350nm by loss type and by sub-system, are presented in Fig.~\ref{fig:throughputall} and \ref{fig:throughput}, respectively.

As a guideline, with this throughput we use the simplified BlueMUSE Exposure Time Calculator (available at \url{https://calc-bluemuse.univ-lyon1.fr/}) and expect a point source to be detected in continuum at 5-sigma (5 x 5 spatial and 1 spectral pixel) down to $\sim$2.9\,10$^{-18}$ erg/s/cm$^2$/\AA in 1 hr and down to $\sim$8.7\,10$^{-19}$ erg/s/cm$^2$/\AA\ at 465 nm in 10 hrs.

\subsection{Image quality}
We have defined an image quality budget for BlueMUSE using a root sum squared method: 
\begin{itemize}
    \item{The image quality requirement is first broken down into static and dynamic errors.}
    \item{Static contributions are allocated with shares representative of MUSE experience (85\% for design and 15\% for MAIT/V). The design contribution is shared among sub-systems according to their relative image quality in the current optical design, with additional contributions accounting for the detector and the pipeline resampling error.}
    \item{Dynamic contributions comprise the effects of instrument stability (both drift and jitter), derotator wobble and relative atmospheric dispersion as described above. When relevant, we estimate contributions assuming a typical exposure time of 30 minutes.}
\end{itemize}

\begin{figure}[ht!]
\begin{minipage}{9cm}
\includegraphics[width=9cm]{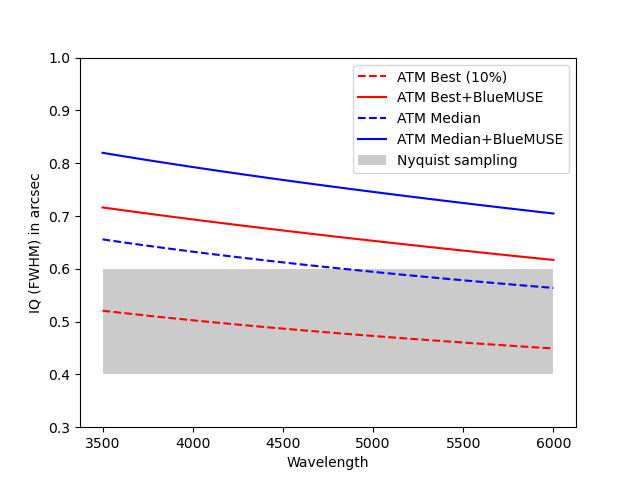}
\end{minipage}
\begin{minipage}{8cm}
 \caption
    { \label{fig:IQ} 
 Global image quality (FWHM of the PSF) expected for BlueMUSE observations as a function of wavelengths, under two cases of atmospheric turbulence (10\% best and median seeing conditions). When adding the contributions from the telescope and instrument we can see that the PSF will always be at least Nyquist sampled by the 0.2"$\times$0.3" rectangular spaxels (grey rectangle).}
\end{minipage}
   
\end{figure}

\begin{figure}[ht!]
\begin{minipage}{4.5cm}
\includegraphics[width=4.5cm]{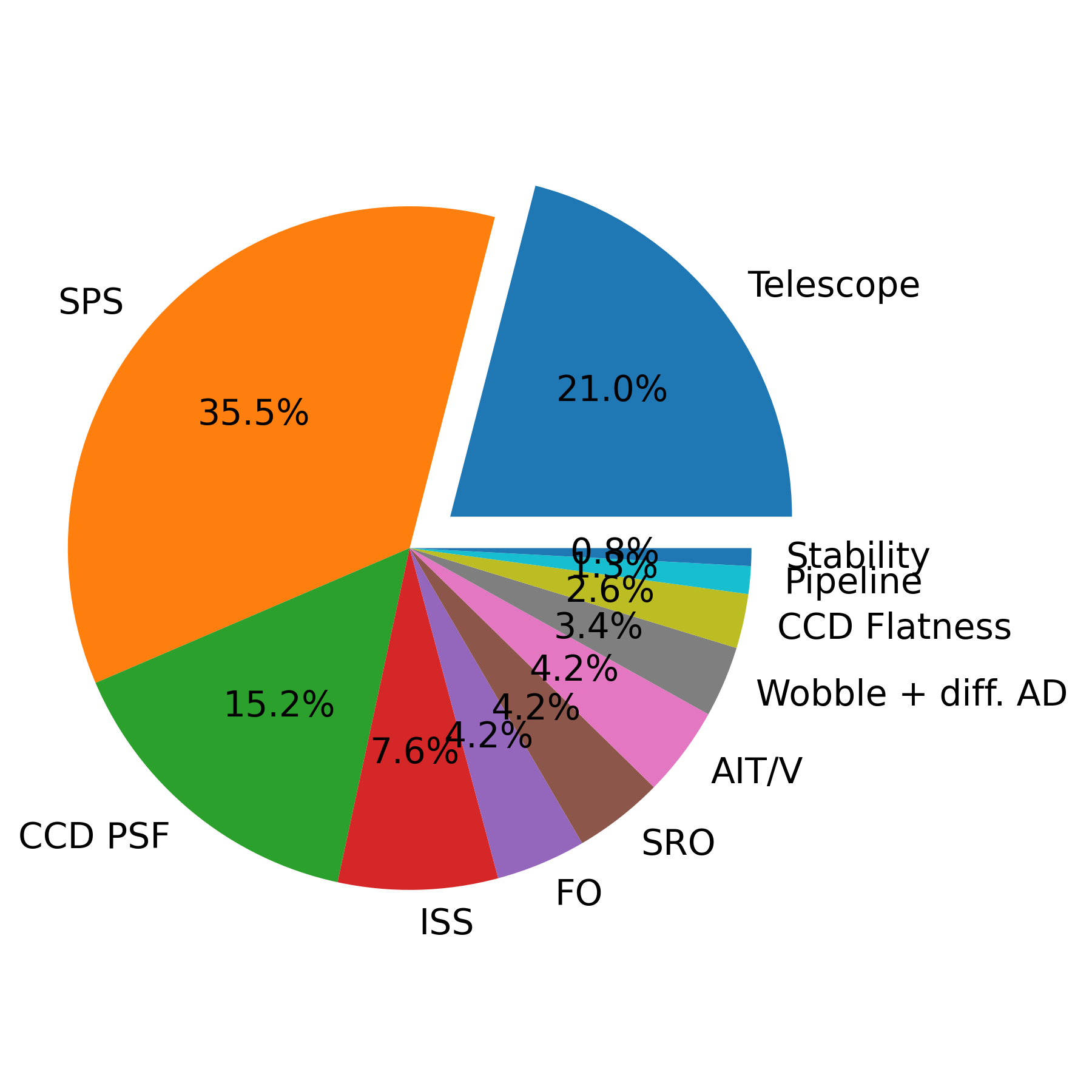}
\end{minipage}
\begin{minipage}{12cm}
\includegraphics[width=12cm]{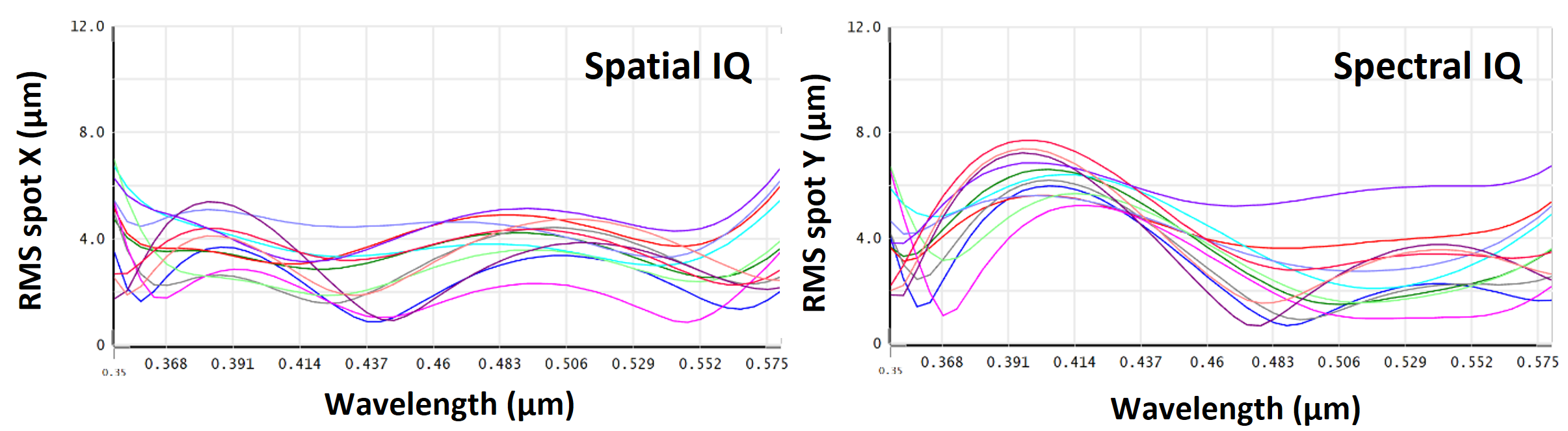}
\end{minipage}
    \caption
    { \label{fig:IQsub} 
 Breakdown of the overall image quality budget by subsystem, which is quite similar at all wavelengths. The contribution from the VLT itself is highlighted for reference. The dominant contribution comes from the BlueMUSE spectrograph, and we show in the right panels the rms spot radius of the spectrograph in both directions (spatial and spectral).}
\end{figure}

The overall image quality budget is presented in Fig.~\ref{fig:IQ} and its breakdown by subsystem is shown in Fig.~\ref{fig:IQsub}.

\section{Conclusion}

We have presented here the main characteristics of the BlueMUSE panoramic IFS in preparation for the VLT, as well as the flow-down from the main science  cases to performance top level requirements. Opening the parameter space at Blue / UV wavelengths and enhanced spectral resolution will open a number of unique science cases from comets in the solar system to massive stars in our galaxy and extragalactic deep fields at cosmic noon.

The first view on the BlueMUSE design architecture and optical design show that we are able to fulfill all top level requirements presented in Sect.~\ref{sec:science}, but also that we expect strong challenges to achieve the throughput and image quality in particular at the limits of blue wavelengths.

The next steps during the Phase A and preliminary design phases will be to flow down these TLRs into instrument technical specifications for each BlueMUSE subsystem, and use them to push further the analysis of the designs, along with a secured plan for the construction phases.


\section*{Acknowledgements}

We acknowledge financial support from the Commission spécialisée Astronomie Astrophysique (CSAA) of CNRS / INSU.
This work is also supported by the Programme National de Cosmologie et Galaxies (PNCG) and the Programme National de Physique Stellaire (PNPS) of CNRS/INSU with INP and IN2P3, co-funded by CEA and CNES. We also acknowledge support from the FRAMA (Fédération de Recherche André-Marie Ampère) as well as the Labex-LIO (Lyon Institute of Origins) under Grant No. ANR-10-LABX-66 (Agence Nationale pour la Recherche).

NC acknowledges funding from the Deutsche Forschungs-gemeinschaft (DFG) - CA 2551/1-1,2. The BlueMUSE project is partially funded through the SNSF FLARE programme for large infrastructures under grants 20FL21\_201777, 20FL20\_216690.
The German contribution to BlueMUSE is funded through the ErUM program of BMBF (project VLT-BlueMUSE-CD, grants 05A23MGA and 05A23BAC).

\FloatBarrier
\bibliography{report} 
\bibliographystyle{spiebib} 

\end{document}